\documentstyle[11pt,newpasp,twoside,epsf]{article}
\def\edcomment#1{\iffalse\marginpar{\raggedright\sl#1\/}\else\relax\fi}
\reversemarginpar

\begin{document}
\title{Horizontal Branch Models as a Test of Mixing on the RGB}
\author{Allen V. Sweigart}
\affil{NASA Goddard Space Flight Center, Code 681, Greenbelt, MD 20771}

\begin{abstract}
We discuss the impact of mixing and rotation along the
red-giant branch (RGB) on the properties of horizontal-branch
(HB) stars with emphasis on two problems: the nature of the
unexpected blue HB population in the metal-rich globular
clusters (GCs) NGC 6388 and NGC 6441 and the cause of the
low gravities in the blue HB stars.  New
stellar models indicate that the sloped HBs in
NGC 6388 and NGC 6441 might arise from a spread in metallicity,
implying that these GCs may be metal-rich analogues of
$\omega$ Cen.  The low gravity problem
can be largely explained by the radiative levitation of Fe in the
atmospheres of the blue HB stars. We show
that the onset of radiative levitation and the drop
in HB rotation velocities at $T_{\rm eff} \, \approx \, 11,000$ K
coincide with the disappearance of surface convection. The low
rotation velocities of the hotter HB stars may be due to the
spin down of the surface layers by a weak stellar wind induced
by the radiative levitation of Fe.  We
conclude that the impact of mixing and rotation on the HB
remains to be clearly established.
\end{abstract}

\section{Introduction}

The abundance anomalies in C, N, O, Na and Al found in red-giant
stars in GCs are often attributed in part to the rotationally
driven mixing of nucleosynthesized material from the vicinity
of the hydrogen shell out to the stellar surface (Kraft 1994). It
has been frequently suggested that such mixing and rotation might
impact the subsequent HB evolution particularly in regard
to the 2nd parameter effect.  Perhaps the best evidence
for this possibility comes from the 2nd parameter clusters
M3 and M13.  In M13 one finds evidence for extensive mixing on the
RGB, high HB rotation velocities and a very blue HB morphology,
while in M3 there is less mixing on the RGB, lower HB rotation
velocities and a redder HB morphology.
            
How would rotation and mixing affect a star's evolution?  Since rotation
delays the helium flash at the tip of the RGB, it would
lead to a larger helium-core mass and greater mass loss.  Consequently
a rotating star would be both bluer and brighter on the HB than
its non-rotating counterpart.  In the case of mixing one must
distinguish between ``shallow'' mixing, which does not penetrate
into the hydrogen shell, and ``deep'' mixing, which
does.  ``Shallow'' mixing would have little, if any, effect on the
HB.  In contrast, ``deep'' mixing into the shell
would increase the envelope helium
abundance as well as the luminosity (and hence mass loss)
at the tip of the RGB.  The net effect of both rotation
and ``deep'' mixing is to produce a bluer and brighter HB morphology.

\vspace{\fill}
\begin{figure*}[t]
\hspace{0.08in}{\epsfbox{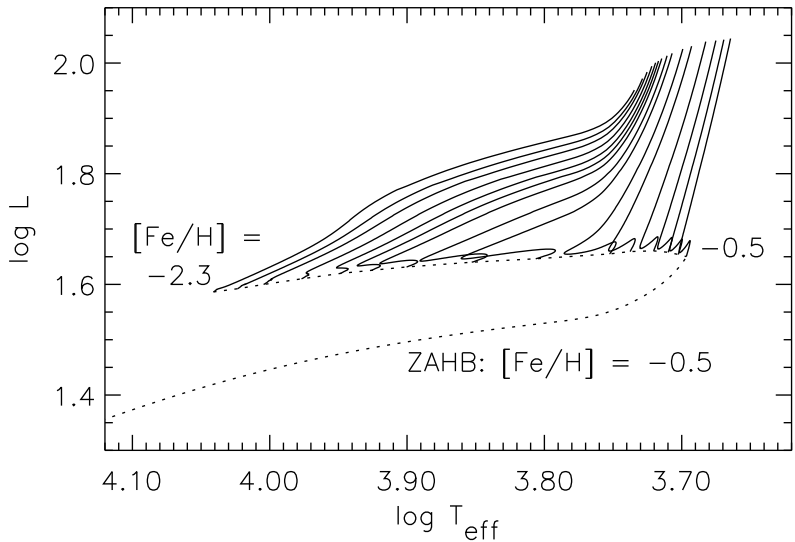}}
\vspace{-0.10in}
\caption{HB tracks (solid curves) for a range in [Fe/H] from $-$0.5
to $-$2.3 and for a mass loss parameter
$\eta_{\rm R}$ = 0.5.  The lower dotted curve represents
the canonical ZAHB for a constant [Fe/H] = $-$0.5.}
\label{figure 1}
\end{figure*}
\vspace{-0.30in}
\section{New Results}

Rich et al. (1997) discovered that the HBs in NGC 6388 and NGC 6441
possess prominent blue extensions not found in other metal-rich
GCs. Quite remarkably, these HBs also slope upward
with decreasing {\bv} from the red clump to the top of the
blue tail.  Using canonical HB models, Sweigart \& Catelan
(1998, hereafter SC98) demonstrated that differences in
age or mass loss along the RGB - two popular 2nd parameter
candidates - cannot explain the HB morphology of these
GCs.  However, SC98 did find that
deep mixing and rotation could produce upward
sloping HBs similar to those observed in these clusters.
                
One scenario not considered by SC98 was a spread in metallicity
within NGC 6388 and NGC 6441, a possibility first raised by
Piotto et al. (1997). In order to explore this scenario in detail,
we have evolved sequences from the main sequence through the
HB phase for a range in [Fe/H] from $-$0.5 (the approximate
metallicity of NGC 6388 and NGC 6441) to $-$2.3.  For simplicity
we assume that all models are coeval and that the Reimers mass loss
parameter $\eta_{\rm R}$ is the same for all [Fe/H]. As
shown in Figure 1, the zero-age HB (ZAHB) for these variable
metallicity tracks is about 0.4 mag brighter than the
canonical ZAHB for [Fe/H] = $-$0.5 at the top of the blue
tail.  When translated into the observational ($V$, {\bv}) plane,
we would expect the HB defined by the tracks
in Figure 1 to slope upward by about the amount observed
in NGC 6388 and NGC 6441. These results suggest that NGC 6388 and
NGC 6441 might be metal-rich analogues of $\omega$ Cen, the only
other GC known to contain a spread in metallicity.

Blue HB stars in the
temperature range $4.3 \, \ga \, {\rm log} \, T_{\rm eff} \, \ga \, 4.0$
generally have lower surface gravities than predicted by canonical
models when their spectra are analyzed using stellar atmospheres with
the cluster metallicity. Deep mixing was initially suggested as a
possible explanation for this low gravity offset. However, it is now
known that radiative levitation of heavy elements can increase
[Fe/H] in the atmospheres of these stars to solar or super solar
values. Most of the low gravity problem disappears when these stars
are analyzed with metal-rich atmospheres (Moehler et al. 2000).
                 
A number of interesting phenomena occur in HB stars around
a temperature of ${\approx} 11,000$ K including the onset of
radiative levitation, a shift to lower surface gravities, a drop
in rotation velocities, a jump in the Str\"omgren u magnitudes,
and in some GCs a gap in the HB distribution.  These phenomena
may be related to the disappearance of surface convection and hence
to the formation of a more stable stellar atmosphere. As illustrated in
Figure 2, HB stars cooler than log $T_{\rm eff} \, \approx \, 3.8$
\begin{figure*}[t]
\hspace{0.15in}{\epsfbox{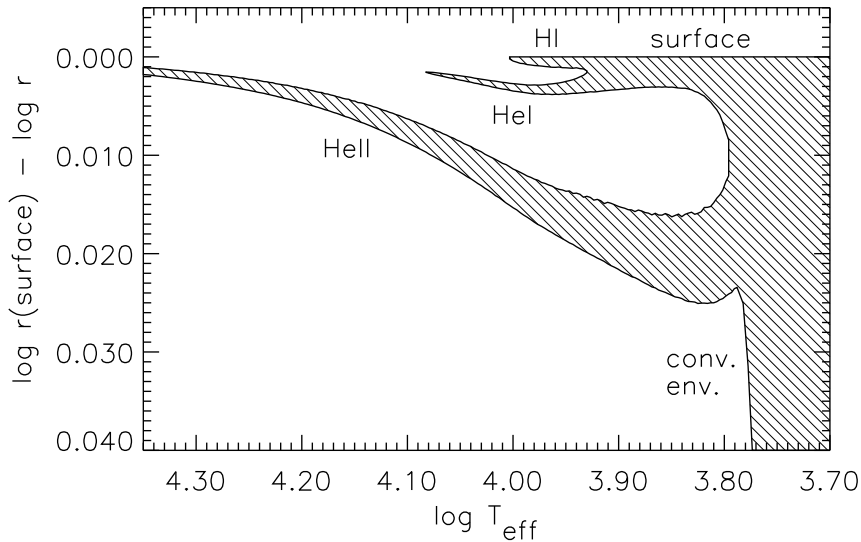}}
\vspace{-0.10in}
\caption{Variation of the envelope convection with $T_{\rm eff}$
along a canonical ZAHB for [Fe/H] = $-$1.6.  Shaded areas
are convective.  The ordinate gives the depth in radius r into
the models.}
\label{figure 2}
\end{figure*}
have deep convective envelopes.  Hotter than this temperature,
the envelope convection breaks into distinct shells associated
with the ionization of H and He. Note that the surface
convection disappears at ${\approx} 11,000$ K.

\end{document}